\documentclass[conference]{IEEEtran}
\usepackage{graphicx}
\usepackage{color}
\usepackage{float}
\usepackage{amsmath,amssymb,amsfonts}
\usepackage{chngpage}
\usepackage{array}
\usepackage{algorithm}
\usepackage{algpseudocode}
\usepackage{tabulary}
\usepackage{booktabs}
\usepackage{cases}
\usepackage{caption}
\usepackage[caption=false,font=footnotesize]{subfig}

\newtheorem{thm} {Theorem}

\newtheorem{cor} {Corollary}

\makeatletter
\floatname{algorithm}{Procedure}

\newcommand{\Rmnum}[1]{\expandafter\@slowromancap\romannumeral #1@}
\makeatother
\begin{document}
\title{Error-Resilient Multicasting for Multi-View 3D Videos in Wireless Networks}
\author{\IEEEauthorblockN{Chi-Heng Lin\IEEEauthorrefmark{1},~De-Nian Yang\IEEEauthorrefmark{1},~Ji-Tang Lee\IEEEauthorrefmark{2} and~Wanjiun Liao\IEEEauthorrefmark{2}}\\
\IEEEauthorrefmark{1}Academia Sinica, Taiwan,
\IEEEauthorrefmark{2}National Taiwan University, Taiwan\\
Email: \IEEEauthorrefmark{1}\{toptoptop3344@gmail.com, dnyang@iis.sinica.edu.tw\},
\IEEEauthorrefmark{2}\{r03942049, wjliao\}@ntu.edu.tw}

\maketitle

\begin{abstract}
With the emergence of naked-eye 3D mobile devices, mobile 3D video services are becoming increasingly important for video service providers, such as Youtube and Netflix, while multi-view 3D videos have the potential to inspire a variety of innovative applications. However, enabling multi-view 3D video services may overwhelm WiFi networks when every view of a video are multicasted. In this paper, therefore, we propose to incorporate depth-image-based rendering (DIBR), which allows each mobile client to synthesize the desired view from nearby left and right views, in order to effectively reduce the bandwidth consumption. Moreover, when each client suffers from packet losses, retransmissions incur additional bandwidth consumption and excess delay, which in turn undermines the quality of experience in video applications. To address the above issue, we first discover the merit of view protection via DIBR for multi-view video multicast using a mathematical analysis and then design a new protocol, named Multi-View Group Management Protocol (MVGMP), to support the dynamic join and leave of users and the change of desired views. The simulation results demonstrate that our protocol effectively reduces bandwidth consumption and increases the probability for each client to successfully playback the desired views in a multi-view 3D video.
\end{abstract}


\section{Introduction}

The \IEEEPARstart{I}{EEE} 802.11 \cite{Stand2012} WiFi standard has achieved
massive market penetration due to its low cost, easy deployment and high
bandwidth. Also, with the recent emergence of naked-eye 3D mobile devices,
such as Amazon's 3D Fire Phone, HTC's EVO 3D, LG's Optimus 3D, and Sharp's
Lynx, mobile 3D video services are expected to become increasingly important
for video service providers such as Youtube and Netflix. In contrast to
traditional stereo single-view 3D video formats, multi-view 3D videos
provide users with a choice of viewing angles and thus are expected to
stimulate the development of innovative applications in television, movies,
education, and advertising \cite{Signal2009}.
Previous research on the deployment of 3D videos in wireless networks has
mostly focused on improving 3D video quality for single-view 3D videos \cite%
{ICC2013, ICCF2013}, but multi-view 3D videos, which
typically offer 5, 16 and 32 different viewing angels \cite{Meet2008} have attracted much less attention. 

Multi-View 3D videos are expected to significantly increase the network load when all views are transmitted.
One promising way to remedy the bandwidth issue is to exploit
depth-image-based rendering (DIBR) in mobile clients, in which the idea is to synthesize the desired view from one left view and one right view \cite{Signal2009}, because 
adjacent left and right views with a sufficiently small angle usually share many similar scenes and objects.
Several schemes for bit
allocation between the texture and depth map \cite{Multimedia2012} and rate control with layered encoding for a multi-view 3D video \cite{ACMMultimedia2012} have been proposed to ensure that
the quality of the synthesized view is very close to the original view
(i.e., by minimizing total distortion or maximizing quality). Therefore,
exploiting DIBR in clients eliminates the need to deliver every view
of a video in a network. For practical situations, the computation overhead and extra energy consumption incurred by DIBR is small enough to be supported by current mobile devices
\cite{ACMMultimedia2012, Processing2012}.For HTTP video streaming (ex., Youtube and MPEG-DASH) with TCP \cite{Dash20, SIGCOMM2011}, instead of UDP, DIBR can be performed when the views are waiting in the streaming buffer before playback.

Equipped with DIBR, only a subset of views are required to be multicasted in a network. However, multi-view 3D video multicast with DIBR brings new challenges in \textit{view selection} for
WiFi networks due to view synthesis and wireless erasure. Firstly, 
the number of skipped views between the left and right views in DIBR 
needs to be constrained to ensure the quality of the synthesized view
\cite{Signal2009}. In other words, since each transmitted view is multicasted to
multiple clients, it is crucial to carefully select the transmitted views so that
the desired view of each user can be synthesized with a left view and a right view close to each other. DIBR has
a quality constraint \cite{Signal2009}, which specifies that the left and
right views are allowed to be at most $R$ views away (i.e., $R-1$ views skipped
between them) to ensure that every view between the left and right view can
be successfully synthesized with good quality. Therefore, each new user
cannot arbitrarily choose a left and a right view for synthesis with DIBR.
The second challenge is that WiFi networks frequently suffer from wireless erasure, 
and different clients suffer from different loss probabilities due to varying channel
conditions \cite{INFOCOM2007, IWCMC2014, IICSP2007}. In 2D and
single-view 3D videos, the \textit{view loss probability} for each user can
be easily derived according to the selected
bit-rate, channel, and the setting of MIMO (e.g., antennas, spatial streams)
in 802.11 networks. For multi-view 3D videos, however, when a video frame is
lost for a user $i$ subscribing a view $k_{i}$, we observe that the left and right views
multicasted in the network to other users can natively serve to \textit{%
protect} view $k_{i}$, since the user $i$ can synthesize the desired view
from the two views using DIBR. However, the view synthesis will fail if only
one left view or one right view is received successfully by the client.
Therefore, a new research problem is to derive the \textit{view failure
probability}, which is the probability that each user does not successfully
receive and synthesize his/her desired view.

In this paper, we first analyze the view 
failure probability and compare it with the traditional view 
loss probability, which is the probability that a view fails to be sent to a user without DIBR. We then propose the Multi-View Group Management Protocol (MVGMP) 
for multi-view 3D multicast. When a user joins the video multicast group, it can 
exploit our analytical result to request the access point(AP) to transmit the most suitable right 
and left views, so that the view failure probability is guaranteed to stay below 
a threshold. On the other hand, when a user leaves the video multicast group, 
the proposed protocol carefully selects and withdraws a set of delivered views to 
reduce the network load, so that the video failure probability for other users will 
not exceed the threshold. Bandwidth consumption can be effectively reduced since it 
is not necessary to deliver all the views subscribed by the clients.

The rest of the paper is organized as follows. Section II describes the
system model. Section III analyzes the view loss probability and view
failure probability. Section IV presents the proposed protocol. Section V
shows the simulation results, and Section VI concludes this paper.

\section{System Model}
This paper considers single-cell video multicast in IEEE 802.11 networks, where the views transmitted by different bit-rates and on different channels are associated with different loss probabilities \cite{INFOCOM2007, IWCMC2014, IICSP2007}. Currently, many video services, such as Youtube and Netflix, require reliable transmissions since Flash or MPEG DASH \cite{Dash20} are exploited for video streaming. Nevertheless, the current IEEE 802.2 LLC protocol for IEEE 802.11 networks does not support reliable multicast transmissions \cite{Stand1998}, and error recovery therefore needs to be handled by Layer-3 reliable multicast standards, such as PGM \cite{PGM2001}.

A multi-view 3D video can be encoded by various encoding schemes \cite{Circuits2007, Proceed2011}. Each view in a video consists of a texture image and a depth map of the corresponding viewing angle. The idea of DIBR is to synthesize a view according to its neighbor left view and neighbor right view. Since the angle between the neighbor left and right views is relatively small, it is expected that the video objects in the synthesized view can be warped (i.e., bent) from those in the two neighbor views. Effective techniques in computer vision and image processing have been proposed to ensure the video quality and limit the processing delay \cite{Processing2011}. 

For example, suppose there are three multicast views, i.e., view 1, 3, and 4 subscribed by all clients. In the original WiFi multicast without DIBR, AP separately delivers each view in a multicast group to the corresponding clients, and three views are separately recovered or retransmitted during packet losses. In contrast, our approach enables a subscribed view to be synthesized by neighbor left and right views with DIBR, while the quality constraint in DIBR states that there are at most $R-1$ views between the neighbor left and right views, and $R$ can be set according to \cite{Signal2009}. When $R=3$ in the above example, the lost of view 3 can be recovered by view 1 and 4, since view 3 can be synthesized by view 1 and 4 accordingly. In other words, a user can first try to synthesize the view according to the left view and right view when a subscribed view is lost, by joining the multicast groups corresponding to the left and right views.

The intuition behind our idea is \textit{traffic protection} from neighbor views. A user can join more multicast groups to protect the desired view without extra bandwidth consumption in the network, because the nearby left view and right view may be originally multicasted to other users that subscribe the views. However, more unnecessary traffic will be received if the desired view is not lost, and the trade-off will be explored in the next section.

\section{Analytical Solution\label{sec: analysis}}
In this section, we present the analytical results for multi-view 3D multicast in multi-rate multi-channel IEEE 802.11 networks with DIBR. We first study the scenario of
single-view subscription for each user and then extend it to multi-view
subscription. Table \uppercase\expandafter{\romannumeral1} summarizes the
notations in the analysis. Based on the mathematical analysis, a new
protocol is proposed in the next section to dynamically assign the proper
views to each user.

\subsection{Single View Subscription}
In single-view subscription, each user $i$ specifies only one desired view $%
k_{i}$. Each view can be sent once or multiple times if necessary. Let $%
p_{i,c,r}$ represent the \textit{view loss probability}, which is the
probability that user $i$ does not successfully receive a view under channel $c$ and bit-rate $r$. 
We define a new probability $P_{\varepsilon }^{(i)}$ for multi-view 3D videos, called \textit{view
failure probability}, which is the probability that user $i$ fails to
receive and synthesize the desired view because the view and nearby
left and right views for synthesis are all lost. In other words,
the view loss probability considers only one view, while the view failure
probability jointly examines the loss events of multiple views.

\begin{thm}
In the theorem, we first explore the most generalized case studied in \cite{INFOCOM2005,MobiCom2005,Transac2009} with each multi-radio client able to operate on multiple channels simultaneously.  
For single-view subscription, the view failure probability for user $i$ is
\begin{align}
& P_{\varepsilon }^{(i)}=\prod_{c\in C_{i},r\in
D_i}p_{i,c,r}^{n_{k_{i},c,r}}\times 
\Bigg(\mathbf{1}\{k_{i}=1\}+\mathbf{1}\{k_{i}=M\}\Bigg) \notag  \label{formula1} \\
& + \sum_{k=1}^{R-1}\Bigg((1-\prod_{c^{\prime }\in C_{i},r^{\prime }\in
D_i}p_{i,c^{\prime },r^{\prime }}^{n_{k_{i}-k,c^{\prime },r^{\prime
}}})\prod_{l=1}^{\min (R-k,M-k_{i})}\prod_{\underset{c_{1},c_{2}\in C_{i}}{%
r_{1},r_{2}\in D_i}}  \notag \\
&\prod_{q=0}^{k-1}p_{i,c_{1},r_{1}}^{n_{k_{i}-q,c_{1},r_{1}}}p_{i,c_{2},r_{2}}^{n_{k_{i}+l,c_{2},r_{2}}}%
\mathbf{1}\{M-1\geq k_{i}\geq k+1\}\Bigg)  \notag \\
&+\prod_{q=0}^{\min (R-1,k_{i}-1)}\prod_{c_{3}\in C_{i},r_{3}\in
D_i}p_{i,c_{3},r_{3}}^{n_{k_{i}-q,c_{3},r_{3}}}\mathbf{1}\{M-1\geq k_{i}\geq
2\}\notag  
\end{align}%
where $\mathbf{1}\{\cdot \}$ denotes the indicator function.
\end{thm}

\textit{Proof:} The view failure event occurs when both of the following two
conditions hold: 1) user $i$ does not successfully receive the desired
view, and 2) user $i$ fails to receive any feasible set consisting of a left view and a
right view with the view distance at most $R$ to synthesize the desired
view. The probability of the first condition is $\prod_{c\in C_{i},r\in
D_i}p_{i,c,r}^{n_{k_{i},c,r}}$ when the the desired view $k_{i} $ of user $i$
is transmitted by $n_{k_{i}}$ times. Note that if the desired view of user $i
$ is view $1$ or view $M$, i.e., $k_{i}=1$ or $k_{i}=M $, user $i$ is not
able to synthesize the desired view with DIBR, and thus the view failure
probability can be directly specified by the first condition. For every
other user $i$ with $M-1\geq k_{i}\geq 2$, we define a set of
non-overlapping events $\{\mathcal{B}_{k}\}_{k=0}^{R-1}$, where $\mathcal{B}%
_{k}$ with $k>0$ is the event that the nearest left view received by user $i$
is $k_{i}-k$ , but user $i$ fails to receive a feasible right view to
synthesize the desired view. On the other hand, $\mathcal{B}_{0}$ is the
event that the user $i$ fails to receive any left view. Therefore, $%
\bigcup_{k=0}^{R-1}\mathcal{B}_{k}$ jointly describes all events for the
second condition.

\begin{table}[t]
\caption{Notations in Analysis.}
\label{table1}
\begin{center}
\begin{tabular}{|l|l|}
\hline
\textbf{Description} & \textbf{Notation} \\ \hline
$R$ & Quality constraint of DIBR \\ \hline
$M$ & Total number of views  \\ \hline
$k_i$ & The view desired by user $i$ \\ \hline
$D_i$ & A set of the available data rates for user $i$ \\ \hline
$C_i$ & A set of the available channels for user $i$ \\ \hline
$n_{j,c,r}$ & Number of multicast transmissions for view $j$ \\
& transmitted by rate $r$ in the channel $c$ \\ \hline
$p_{i,c,r}$ & The view loss probability for user $i$ under \\
& channel $c$ and rate $r$ \\ \hline
$P_{\varepsilon}^{(i)}$ & The probability that user $i$ cannot obtain the \\
& desired view either by direct transmission or by \\
& DIBR \\ \hline
$p_{c,r}^{\text{AP}}(n)$ & The probability that AP multicasts a view $n$
times \\
& under the channel $c$ and the rate $r$ \\ \hline
$\alpha_i$ & The percentage of the desired views that can be \\
& received or synthesized successfully by user $i$ \\
\hline
$p_{\text{select}}$ & The probability that a user selects each view \\
\hline
\end{tabular}%
\end{center}
\end{table}

For each event $\mathcal{B}_{k}$ with $k>0$,
\begin{align}
& P(\mathcal{B}_{k})=(1-\prod_{c^{\prime }\in C_{i},r^{\prime }\in
D_i}p_{i,c^{\prime },r^{\prime }}^{n_{k_{i}-k,c^{\prime },r^{\prime
}}})\prod_{l=1}^{\min (R-k,M-k_{i})}  \notag \\
&\prod_{\underset{c_{1},c_{2}\in C_{i}}{r_{1},r_{2}\in D_i}%
}\prod_{q=0}^{k-1}p_{i,c_{1},r_{1}}^{n_{k_{i}-q,c_{1},r_{1}}}p_{i,c_{2},r_{2}}^{n_{k_{i}+l,c_{2},r_{2}}}%
\mathbf{1}\{M-1\geq k_{i}\geq k+1\}  \notag
\end{align}%
The first term $1-\prod_{c^{\prime }\in C_{i},r^{\prime }\in
D_i}p_{i,c^{\prime },r^{\prime }}^{n_{k_{i}-k,c^{\prime },r^{\prime }}}$
indicates that user $i$ successfully receives view $k_{i}-k $, and the
second term
\begin{equation*}
\prod_{l=1}^{\min (R-k,M-k_{i})}\prod_{\underset{c_{1},c_{2}\in C_{i}}{%
r_{1},r_{2}\in D_i}%
}\prod_{q=0}^{k-1}p_{i,c_{1},r_{1}}^{n_{k_{i}-q,c_{1},r_{1}}}p_{i,c_{2},r_{2}}^{n_{k_{i}+l,c_{2},r_{2}}}
\end{equation*}%
means that user $i$ does not successfully receive any left view between $%
k_{i}-k$ and $k$ and any right view from $k_{i}+1$ to $k_{i}+\min
(R-k,M-k_{i})$. It is necessary to include an indicator function in the last
term since $\mathcal{B}_{k}$ will be a null event if $k_{i}\leq k$, i.e.,
user $i$ successfully receives a view outside the view boundary. Finally,
the event $\mathcal{B}_{0}$ occurs when no left view is successfully received
by user $i$.
\begin{align}
&P(\mathcal{B}_{0})=  \notag \\
&\prod_{q=0}^{\min (R-1,k_{i}-1)}\prod_{c_{3}\in C_{i},r_{3}\in
D_i}p_{i,c_{3},r_{3}}^{n_{k_{i}-q,c_{3},r_{3}}}\mathbf{1}\{M-1\geq k_{i}\geq
2\}  \notag
\end{align}%
The theorem follows after summarizing all events. $\blacksquare $ \newline

\textbf{Remark:}  The advantage of a multi-view 3D multicast with DIBR 
can be clearly seen when comparing the view loss probability and view 
failure probability. The latter probability attaches a new term (i.e., the
probability of $\bigcup_{k=0}^{R-1}\mathcal{B}_{k}$) to the view loss probability, where a larger $R$ reduces the probability of the second term. Equipped with DIBR, therefore, the view failure probability is much smaller than the view loss probability, see Section \ref{sec: simulation}.

For the case that each single-radio client can access only one channel and rate at any time, the theorem can be changed to the following one.\footnote{$\displaystyle P_{\varepsilon }^{(i)}=\prod_{r\in D_i}p_{i,c,r}^{n_{k_{i},c,r}}\times \Bigg(\mathbf{1}\{k_{i}=1\}+\mathbf{1}\{k_{i}=M\}\Bigg)\\+\sum_{k=1}^{R-1}\Bigg((1-\prod_{r^{\prime }\in D_i}p_{i,c,r^{\prime }}^{n_{k_{i}-k,c,r^{\prime}}})\prod_{l=1}^{\min (R-k,M-k_{i})}\prod_{r_{1},r_{2}\in D_i}\prod_{q=0}^{k-1}\\p_{i,c,r_{1}}^{n_{k_{i}-q,c,r_{1}}}p_{i,c,r_{2}}^{n_{k_{i}+l,c,r_{2}}}\mathbf{1}\{M-1\geq k_{i}\geq k+1\}\Bigg)\\+\prod_{q=0}^{\min (R-1,k_{i}-1)}\prod_{r_{3}\in D_i}p_{i,c,r_{3}}^{n_{k_{i}-q,c,r_{3}}}\mathbf{1}\{M-1\geq k_{i}\geq 2\}$}

\subsection{Multiple View Subscription}

In the following, we explore the case of a user desiring to subscribe to multiple 
views. We first study the following two scenarios: 1) every view is
multicasted; 2) only one view is delivered for every $\widetilde{R}$ views, $%
\widetilde{R}$ $\leq R$, and thus it is necessary for a user to synthesize
other views accordingly. We first define $\alpha _{i}$, which represents the
percentage of desired views that can be successfully received or synthesized by 
user $i$.
\begin{equation*}
\alpha _{i}=\frac{\sum_{k_{i}\in \mathcal{K}_{i}}\mathbf{1}\{\text{user }i%
\text{ can obtain view }k_{i}\}}{|\mathcal{K}_{i}|}
\end{equation*}%
where $\mathcal{K}_{i}$ denotes the set of desired views for user $i$. By using Theorem 1, we can immediately arrive at the following corollary.

\begin{cor}
\begin{align}  \label{formula2}
\mathbb{E}[\alpha_i]=\frac{\sum_{k_i\in\mathcal{K}_i}(1-P_{
\varepsilon}^{(i)}(k_i))}{|\mathcal{K}_i|}
\end{align}
where $P_{\varepsilon}^{(i)}(k_i)$ is given in Theorem 1.
\end{cor}

\textit{Proof:}
\begin{align}
\mathbb{E}[\alpha _{i}]=& \frac{\sum_{k_{i}\in \mathcal{K}_{i}}\mathbb{E}%
\mathbf{1}\{\text{user }i\text{ can obtain view }k_{i}\}}{|\mathcal{K}_{i}|}
\notag \\
=& \frac{\sum_{k_{i}\in \mathcal{K}_{i}}(1-P_{\varepsilon }^{(i)}(k_{i}))}{|%
\mathcal{K}_{i}|}  \notag
\end{align}%
$\blacksquare $

Eq. (1) becomes more complicated as $|\mathcal{K}_{i}|$ increases. In the
following, therefore, we investigate the asymptotic behavior of $\alpha _{i}$ for a
large $|\mathcal{K}_{i}|$ and a large $M$ (i.e., $|\mathcal{K}_{i}|\leq
M $). To find the closed-form solution, we first consider a uniform view
subscription and assume that user $i$ subscribes to each view $j$ with
probability $p_{\text{select}}=\frac{|\mathcal{K}_i|}{M}$ independently
across all views so that the average number of selected views is $|\mathcal{K%
}_i|$. Assume the AP multicasts view $j$ in channel $c$ with rate $r$ by $n$
times with probability $p_{j,c,r}^{\text{AP}}(n)$ independently across all
views, channels, and rates. In the following, we first perform the asymptotic analysis to derive the theoretical closed-form solution, and we then present the
insights from the theorem by comparing the results of single-view
subscription and multi-view subscription.

\begin{thm}
In multi-view 3D multicast,
\begin{align}
\alpha _{i}(\mathcal{K}_i)\overset{a.s.}{\rightarrow }& (1-p_i)\Bigg\{%
\sum_{k=1}^{R}k(1-p_i)p_i^{k-1}+p_i^{R}\Bigg\} \\
\mathbb{E}[\alpha _{i}(\mathcal{K}_i)]\overset{a.s.}{\rightarrow }& (1-p_i)%
\Bigg\{\sum_{k=1}^{R}k(1-p_i)p_i^{k-1}+p_i^{R}\Bigg\}
\end{align}%
as $|\mathcal{K}_i|\rightarrow \infty $, where $p_i=\prod_{c\in C_{i},r\in
D_i}\sum_{n}p_{c,r}^{\text{AP}}(n)p_{i,c,r}^{n}$
\end{thm}

\textit{Proof:} We first derive the view loss probability for user $i$.
Suppose that the AP multicasts a view $n$ times via channel $c$ and rate $%
r$. The probability that user $i$ cannot successfully receive the view is $%
p_{i,c,r}^{n}$. Because the AP will multicast a view $n$ times via
channel $c $ and rate $r$ with probability $p_{c,r}^{\text{AP}}(n)$, the
probability that user $i$ cannot receive the view via channel $c$ and rate $r
$ is $\sum_{n}p_{c,r}^{\text{AP}}(n)p_{i,c,r}^{n}$. Therefore, the view loss
probability for user $i$ is the multiplication of the view loss
probabilities in all channels and rates, i.e., $\prod_{c\in C_{i},r\in
D_i}\sum_{n}p_{c,r}^{\text{AP}}(n)p_{i,c,r}^{n}$. For simplification, we
denote $p_i$ as the view loss probability for user $i$ in the remainder of
the proof.

Since the multicast order of views is not correlated to $\alpha _{i}$, we
assume that the AP sequentially multicasts the views from view $1$ to view $M$. 
Now the scenario is similar to a tossing game, where we toss $%
M $ coins, and a face-up coin represents a view successfully received from the
AP. Therefore, the face-up probability of at least one coin is $1-p_i^M$. Now we mark a 
coin with probability $p_{\text{select}}$ if it is face-up or if there is one former
tossed face-up coin and one latter tossed face-up coin with the view
distance at most $R$. Since the above analogy captures the mechanism of
direct reception and DIBR of views, the marked coins then indicate that the
views selected by user $i$ can be successfully acquired.

To derive the closed-form asymptotic result, we exploited the \textit{delayed renewal
reward process}, in which a cycle begins when a face-up coin appears, and the
cycle ends when the next face-up coin occurs. The reward is defined as the
total number of marked coins. Specifically, let $\{N(t):=\sup
\{n:\sum_{i=0}^{n}X_{i}\leq t\},t\geq 0\}$ denote the delayed renewal reward
process with inter-arrival time $X_{n}$, where $X_{n}$ with $n\geq 1$ is the
time difference between two consecutive face-up coins, and $X_{0}$ is the
time when the first face-up coin appears.

Let $R(M)$ and $R_{n}$ denote the total reward earned at the time $M$, which
corresponds to the view numbers in a multi-view 3D video. At cycle $n$,
\begin{equation*}
\frac{R(M)}{M}=\frac{\sum_{n=1}^{N(M)}R_{n}}{M}+o(1)~~~a.s.
\end{equation*}%
where the $o(1)$ term comes from the fact that the difference between the total
reward and $\sum_{n=1}^{N(M)}R_{n}$ will have a finite mean. Recall that the
reward earned at each cycle is the number of marked coins,
\begin{numcases}{\mathbb{E}[R_n|X_n]=}
   p_{\textrm{select}},  & for $X_n > R$\nonumber\\
   X_np_{\textrm{select}}, & for $X_n\leq R$
  \end{numcases}
since when $X_{n}\leq R$, $X_{n}$ coins can be marked (each
with probability $p_{\text{select}}$) between two consecutive face-up coins,
and thus the expected reward given $X_n$ is $X_np_{\text{select}}$. By
contrast, only one coin can be marked with probability $p_{\text{select}}$
when $X_{n}>R$, and the expectation of reward given $X_n$ is only $p_{\text{%
select}}$.

Since $X_{n}$ is a geometric random variable with parameter $1-p_i$, we have
\begin{equation*}
\mathbb{E}[X_{n}]=1-p_i+2p_i(1-p_i)+3p_i^{2}(1-p_i)+\cdots =\frac{1}{%
1-p_i}
\end{equation*}%
and
\begin{align}
\mathbb{E}[R_{n}]=& p_{\text{select}}(1-p_i)+2p_{\text{select}%
}p_i(1-p_i)+\cdots  \notag \\
& +Rp_{\text{select}}p_i^{R-1}(1-p_i)+p_{\text{select}}p_i^{R}
\end{align}%
By theorem 3.6.1 of renewal process in \cite{ross},
\begin{align}
\frac{\sum_{n=1}^{N(M)}R_{n}}{M}& \overset{a.s.}{\rightarrow }\frac{\mathbb{E%
}R_{n}}{\mathbb{E}X_{n}}  \notag \\
& =p_{\text{select}}(1-p_i)\Bigg\{\sum_{k=1}^{R}k(1-p_i)p_i^{k-1}+p_i^{R}%
\Bigg\}
\end{align}%
Let $U_{M}$ denote the number of views selected by user $i$. Therefore,
\begin{equation*}
\alpha _{i}=\frac{R(M)}{U_{M}}=\frac{R(M)}{M}\frac{M}{U_{M}}
\end{equation*}%
For $\frac{U_{M}}{M}\overset{a.s.}{\rightarrow }p_{\text{select}}$, by the
strong law of large numbers, after combining with Eq. (4), (5), (6),
\begin{equation*}
\alpha _{i}\overset{a.s.}{\rightarrow }(1-p_i)\Bigg\{%
\sum_{k=1}^{R}k(1-p_i)p_i^{k-1}+p_i^{R}\Bigg\}
\end{equation*}%
The proof for convergence in mean is similar. It is only necessary to
replace the convergence in Eq. (6) by the convergence in mean, which can be proven by the 
same theorem. $\blacksquare $

\textbf{Remark:} Under the above uniform view subscription, it can be observed that $%
\alpha _{i}$ is irrelevant to $p_{\text{select}}$, implying
that different users with different numbers of subscription will acquire the
same percentage of views. Most importantly, $\alpha _{i}=1-p_i$
for multi-view 3D multicasts without DIBR. In contrast, multi-view 3D
multicasting with DIBR effectively improves $\alpha _{i}$ by $%
\sum_{k=1}^{R}k(1-p_i)p_i^{k-1}+p_i^{R}$. Since this term is strictly
monotonically increasing with $R$, we have $%
\sum_{k=1}^{R}k(1-p_i)p_i^{k-1}+p_i^{R}>
\sum_{k=1}^{1}k(1-p_i)p_i^{k-1}+p_i=1$, which implies that the percentage of
obtained views is strictly larger in statistic term s by utilizing the DIBR technique.

In the following, we consider the second case with only one view delivered
for every $\widetilde{R}$ view, where the bandwidth consumption can be
effectively reduced. Note that the following corollary is equivalent to
Theorem 2 when $\widetilde{R}=1$.

\begin{cor}
If the AP only transmits one view with probability $p_{c,r}^{\text{AP}}(n)$
for every $\widetilde{R}$ views,
\begin{align}
\alpha _{i}(\mathcal{K}_i)\overset{a.s.}{\rightarrow }& \frac{(1-p_i)\Bigg\{%
\sum_{k=1}^{\lfloor \frac{R}{\widetilde{R}}\rfloor }\widetilde{R}%
k(1-p_i)p_i^{k-1}+p_i^{\lfloor \frac{R}{\widetilde{R}}\rfloor }\Bigg\}}{%
\widetilde{R}} \\
\mathbb{E}[\alpha _{i}(\mathcal{K}_i)]\rightarrow & \frac{(1-p_i)\Bigg\{%
\sum_{k=1}^{\lfloor \frac{R}{\widetilde{R}}\rfloor }\widetilde{R}%
k(1-p_i)p_i^{k-1}+p_i^{\lfloor \frac{R}{\widetilde{R}}\rfloor }\Bigg\}}{%
\widetilde{R}}
\end{align}%
as $|\mathcal{K}_i|\rightarrow \infty $, where $p_i=\prod_{c\in C_{i},r\in
D_i}\sum_{n}p_{c,r}^{\text{AP}}(n)p_{i,c,r}^{n}$
\end{cor}

\section{Protocol Design}

For a multi-view 3D multicast, each view sent in a channel with a rate is associated with a multicast group. Based on the analytical results in Section \ref{sec: analysis}, 
each client subscribes to a set of views by joining a set of multicast 
group, in order to satisfy the view failure probability. To support the dynamic join and leave of users and the change of the subscribed views, we present a new protocol, named Multi-View Group Management Protocol (MVGMP), which exploits the theoretical results in Section III. The MVGMP protocol extends 
the current IETF Internet standard for multicast group management, the 
IGMP \cite{IGMPRFC}, by adding the view selection feature to the protocol. The IGMP is 
a receiver-oriented protocol, where each user periodically and actively updates 
its joined multicasting groups to the designated router (i.e., the AP in this 
paper). 

In MVGMP, the AP maintains a table, named
\textit{ViewTable}, for each video. The table specifies the current
multicast views and the corresponding bit-rates and channels for each view%
\footnote{%
Note that each view is allowed to be transmitted multiple times in different
channels and rates if necessary, as described in Section \ref{sec: analysis}.%
}, and each multicast view is associated with a multicast address and a set
of users that choose to receive the view. ViewTable is periodically
broadcasted to all users in the WiFi cell. The MVGMP includes two control messages. 
The first message is Join, which contains the address of a new user and 
the corresponding requested view(s), which can be the subscribed views, or the left 
and right views to synthesize the subscribed view. An existing user can also 
exploit this message to update its requested views. The second message is Leave, 
which includes the address of a leaving user and the views that no longer need to 
be received. An existing user can also exploit this message to stop receiving a 
view. Following the design rationale of the IGMP, the MVGMP is also a 
soft-state protocol, which implies that each user is required to periodically send 
the Join message to refresh its chosen views, so that unexpected connection drops 
will not create dangling states in ViewTable.

\textbf{Join. }When a new member decides to join a 3D video multicast
transmission, it first acquires the current ViewTable from the AP.
After this, the user identifies the views to receive according to Theorem
1. Specifically, the client first examines whether ViewTable has included
the subscribed view. If ViewTable does not include the subscribed view, or
if the view loss probability for the subscribed view in the corresponding
channel and bit-rate exceeds the threshold, the user adds a left view and a
right view that lead to the maximal decrement on the view failure
probability. To properly select the views,
The user can search the view combinations exhaustively with the theoretical results in Section III, because R is small and thus only a small number of views nearby to the desired view is necessary to be examined. However, a view cannot be added to a channel without sufficient bandwidth.

When a multi-view 3D video starts, usually the current multicast views in
ViewTable are not sufficient for a new user. In other words, when the view
failure probability still exceeds the threshold after the user selects all
transmitted left and right views within the range $R$ in ViewTable, the user
needs to add the subscribed view to ViewTable with the most suitable channel
and bit-rate to reduce the view failure probability. Also, the left and
right views are required to be chosen again according to the analytic results in 
Section \ref{sec: analysis} to avoid receiving too many
views. After choosing the views to be received, a Join message is sent to
the AP. The message contains the views that the user chooses to receive,
and the AP adds the user to ViewTable accordingly. To avoid receiving
too many views, note that a user can restrict the maximum number of left and right
views that are allowed to be received and exploited for DIBR.

\textbf{Leave and View Re-organization. }On the other hand, when a user
decides to stop subscribing to a multi-view 3D video, it multicasts a Leave
message to the AP and to any other user that receives at least one identical
view $k_{i}$. Different from the Join message, the Leave message is also
delivered to other remaining users in order to minimize the bandwidth
consumption, since each remaining user that receives $k_{i}$ will examine if
there is a chance to switch $k_{i}$ to another view $\overline{k}_{i}$ that
is still transmitted in the network. In this case, the remaining user also
sends a Leave message that includes view $k_{i}$, together with a Join
message that contains view $\overline{k}_{i}$. If a view is no longer
required by any remaining users, the AP stops delivering the view. Therefore,
the MVGMP can effectively reduce the number of multicast views.

\textbf{Discussion. }Note that the MVGMP can support the scenario of a user changing 
the desired view, by first sending a Leave message and then a Join message. 
Similarly, when a user moves, thus changing the channel condition, it will send a 
Join message to receive additional views if the channel condition deteriorates, or 
a Leave message to stop receiving some views if the channel condition 
improves. Moreover, when a user is handed over to a new WiFi cell, it first 
sends a Leave message to the original AP and then a Join message to the new AP. If 
the network connection to a user drops suddenly, the AP removes the 
information corresponding to the user in ViewTable when it does not receive the 
Join message (see soft-state update as explained earlier in this section) for a 
period of time. Therefore, the MVGMP also supports the silent leave of a user from 
a WiFi cell. Moreover, our protocol can be extended to the multi-view subscription 
for each client by replacing Theorem 1 with Theorem 2. The fundamental operations of
Join/Leave/Reorganize remain the same since each view is maintained by a
separate multicast group. 

\section{Simulation Results\label{sec: simulation}}

In the following, we first describe the simulation setting and then compare 
the MVGMP with the current multicast scheme.

\subsection{Simulation Setup}

We evaluate the channel time of the MVGMP in a series of scenarios with NS3 
802.11n package. The channel time of a multicast scheme is the average time 
consumption of a frame in WiFi networks. To the best knowledge of the authors, there 
has been no related work on channel time minimization for multi-view 3D video 
multicast in WiFi networks. For this reason, we compare the MVGMP with the 
original WiFi multicast scheme, in which all desired views are multicasted to the 
users.

We adopt the setting of a real multi-view 3D dataset Book Arrival \cite{Meet2008} and the existing multi-view 3D videos \cite{ICIP2007} with 16 views, where the texture video quantization step is 6.5, and depth map quantization step is 13, and the PSNR of the synthesized views in DIBR is around 37dB \cite{Broadcasting2012}. The video rates for reference texture image and its associated depth map are assigned as $r_t=600$kbps and $r_d=200$kbps, respectively, and thus $r=800$ kbps. The DIBR quality constraint is 
3, $R=3$. The threshold of each user is uniformly distributed in (0, 0.1]. Each user 
randomly chooses one preferred view from three preference
distributions: Uniform, Zipf, and Normal distributions. There is no
specifically hot view in the Uniform distribution. In contrast, the Zipf
distribution, $f(k;s;N)=(\frac{1}{k^{s}})/\sum_{n=1}{N}(\frac{1}{n^{s}})$,
differentiates the desired views, where $k$ is the preference rank of a
view, $s$ is the the exponent characterizing the distribution, and $N$ is
the number of views. The views with smaller ranks are major views and thus
more inclined to be requested. We set $s=1$ and $N=|V|$ in this paper. In
the Normal distribution, central views are accessed with higher probabilities. The 
mean is set as $|V|/2$, and the variance is set as 1 throughout this study.

We simulate a dynamic environment with $50$ client users located randomly
in the range of an AP. After each frame, there is an arrival and
departure of a user with probabilities $\lambda $ and $\mu $, respectively.
In addition, a user changes the desired view with probability $\eta $.
The default probabilities are $\lambda =0.2$, $\mu =0.3$, $\eta =0.4$. TABLE
II summarizes the simulation setting consisting of an 802.11n WiFi network
with a 20MHz channel bandwidth and 13 orthogonal channels. In the following,
we first compare the performance of the MVGMP with the current WiFi multicast
scheme in different scenarios and then compare the analytical and simulation
results.

\begin{table}[t]
\caption{Simulation Settings.}
\label{table1}
\begin{center}
\begin{tabular}{|l|l|}
\hline
\textbf{Parameter} & \textbf{Value} \\ \hline
Carrier Frequency & 5.0 GHz \\ \hline
The unit of Channel Time & $1ms$ \\ \hline
Channel Bandwidth & 20MHz \\ \hline
AP Tx Power & 20.0 dBm \\ \hline
OFDM Data Symbols & 7 \\ \hline
Subcarriers & 52 \\ \hline
Video bit-rate(per view) & 800kbps \\ \hline
Number of Orthogonal Channels & 13 \\ \hline
Transmission Data Rates & \{6.5, 13, 19.5, 26, 39, 52, 58.5, 65\} \\
& Mbps defined in 802.11n spec. \cite{Stand2012} \\ \hline
\end{tabular}
\end{center}
\vspace{-8pt}
\end{table}

The relationship between the setting of $R$ and the video quality has been studied in \cite{Signal2009, Processing2011}. Therefore, due to the space constraint, we focus on the channel time and view failure probability with different $R$ here, and the corresponding video quality can be derived according to \cite{Signal2009, Processing2011}.

\subsection{Scenario: Synthesized Range}

Fig. 1 evaluates the MVGMP with different settings of $R$. Compared with the 
current WiFi multicast, the channel time is effectively reduced in the MVGMP 
as $R$ increases. Nevertheless, it is not necessary to set a large $R$ because 
the improvement becomes marginal as $R$ exceeds 3. Therefore, this finding indicates 
that a small $R$ (i.e., limited quality degradation) is sufficient to effectively 
reduce the channel time in WiFi.

\subsection{Scenario: Number of Views}

Fig. 2 explores the impact of the numbers of views in a video. The channel
time in both schemes increases when the video includes more views, because
more views need to be transmitted. This result shows that MVGMP
consistently outperforms the original WiFi multicast scheme with different
numbers of views in a video.

\subsection{Scenario: Number of Users in Steady State}

Fig. 3 evaluates the channel time with different numbers of users in the
steady state. We set $\lambda =\mu =0.25$, so that the expected number of
users in the network remains the same. The channel time was found to grow as the number
of users increases. Nevertheless, the increment becomes marginal since most
views will appear in \textit{ViewTable}, and thus more users will subscribe to the same views in the video.

\subsection{Scenario: Utilization Factor}
Fig. 4 explores the impact of the network load. Here, we change the
\textit{loading ratio} $\rho :=\frac{\lambda }{\mu }$, i.e., the ratio
between the arrival probability $\lambda $ and departure probability $\mu $. Initially, 
new multicast users continuously join the 3D video stream until the network contains 50 
users. The results indicate that the channel time increased for both multicast schemes.
Nevertheless, the MVGMP effectively reduces at least $40\%$ of channel time for all 
three distributions.
\begin{figure*}[!t]
\captionsetup[subfigure]{labelformat=empty}
\centering
\subfloat[Fig. 1: Synthesis Range]{\includegraphics[width=2.2in]{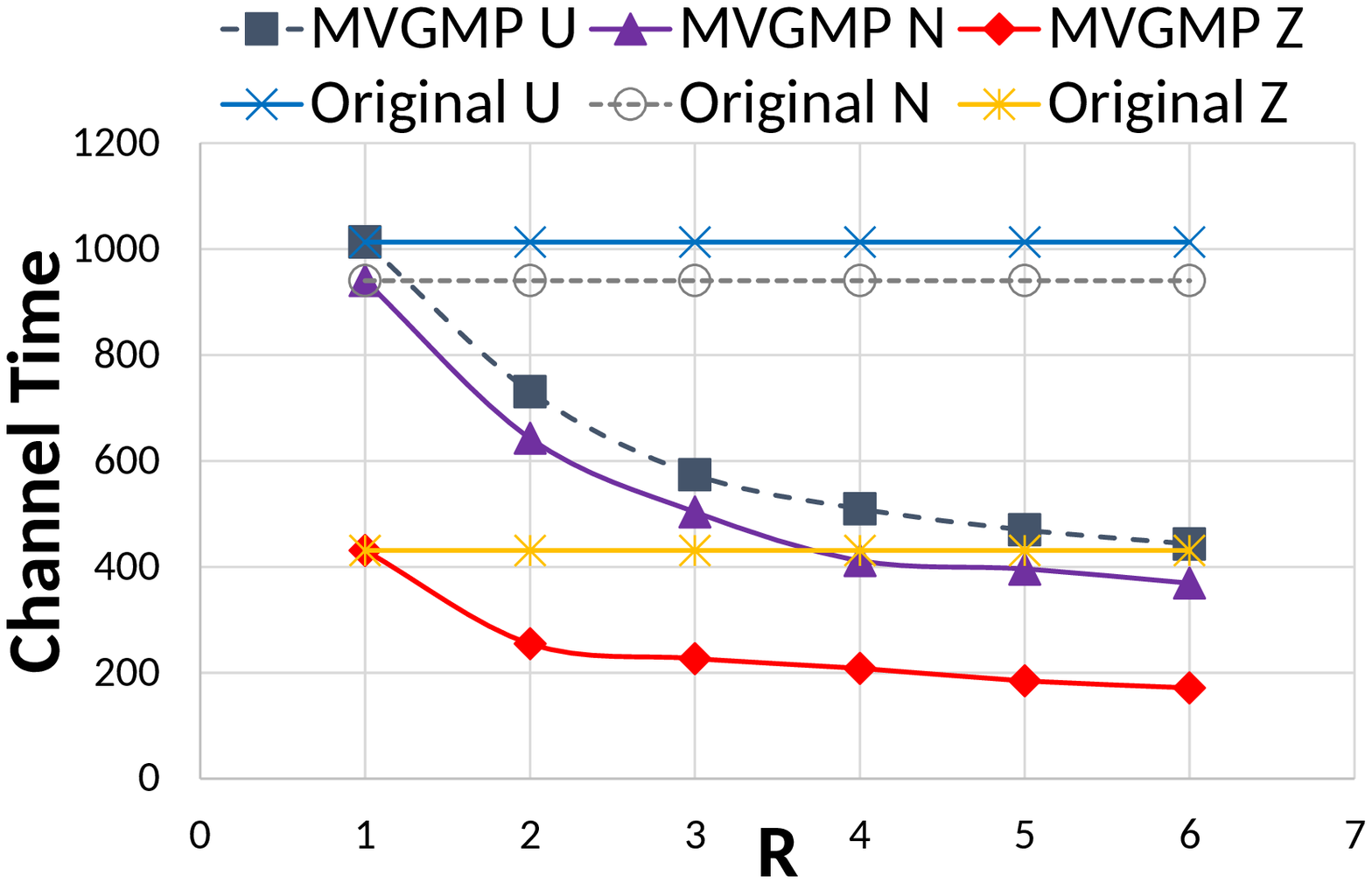}}
\hfil
\subfloat[Fig. 2: Number of Views in a Video]{\includegraphics[width=2.2in]{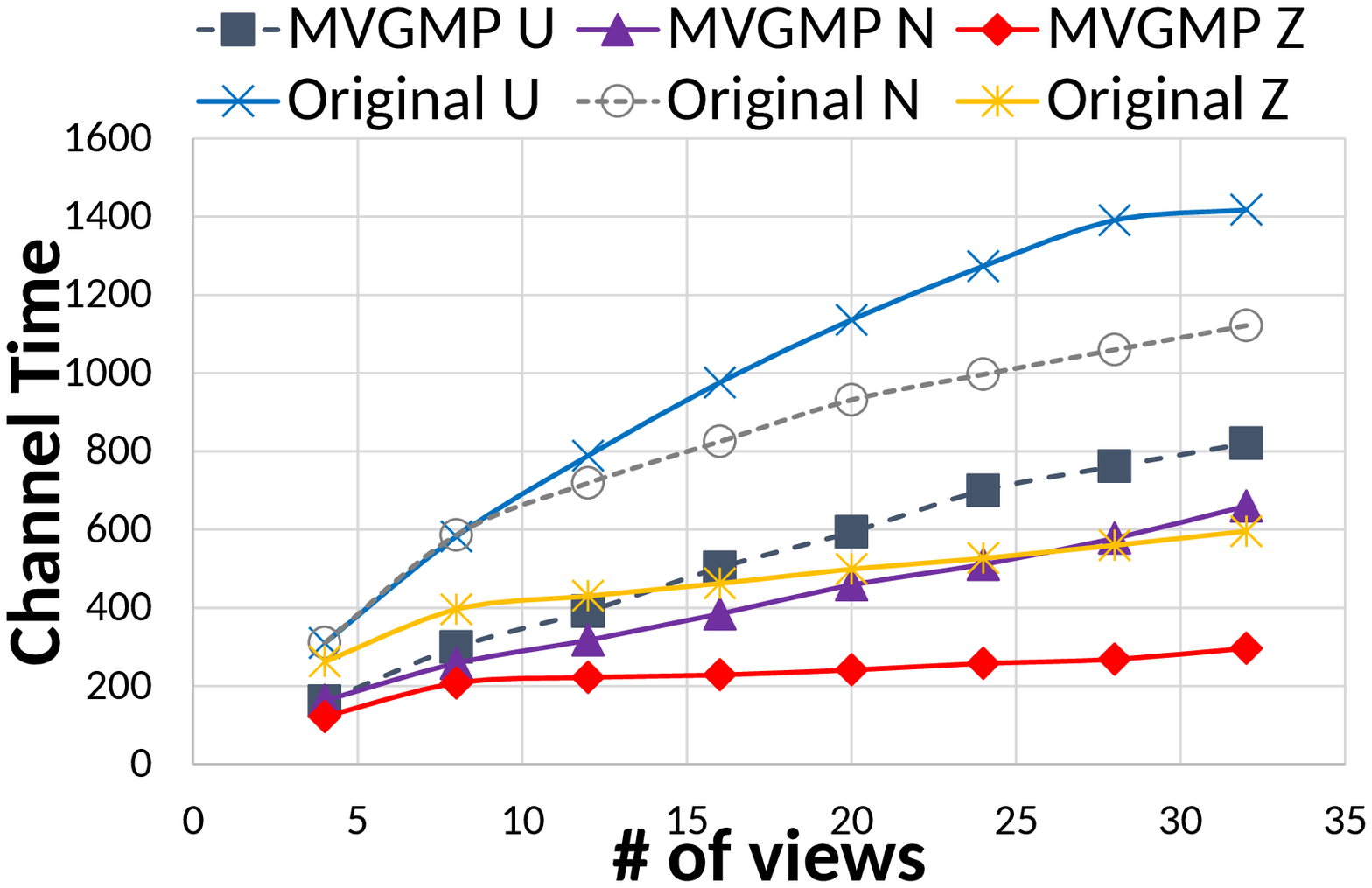}}
\hfil
\subfloat[Fig. 3: Number of Users]{\includegraphics[width=2.2in]{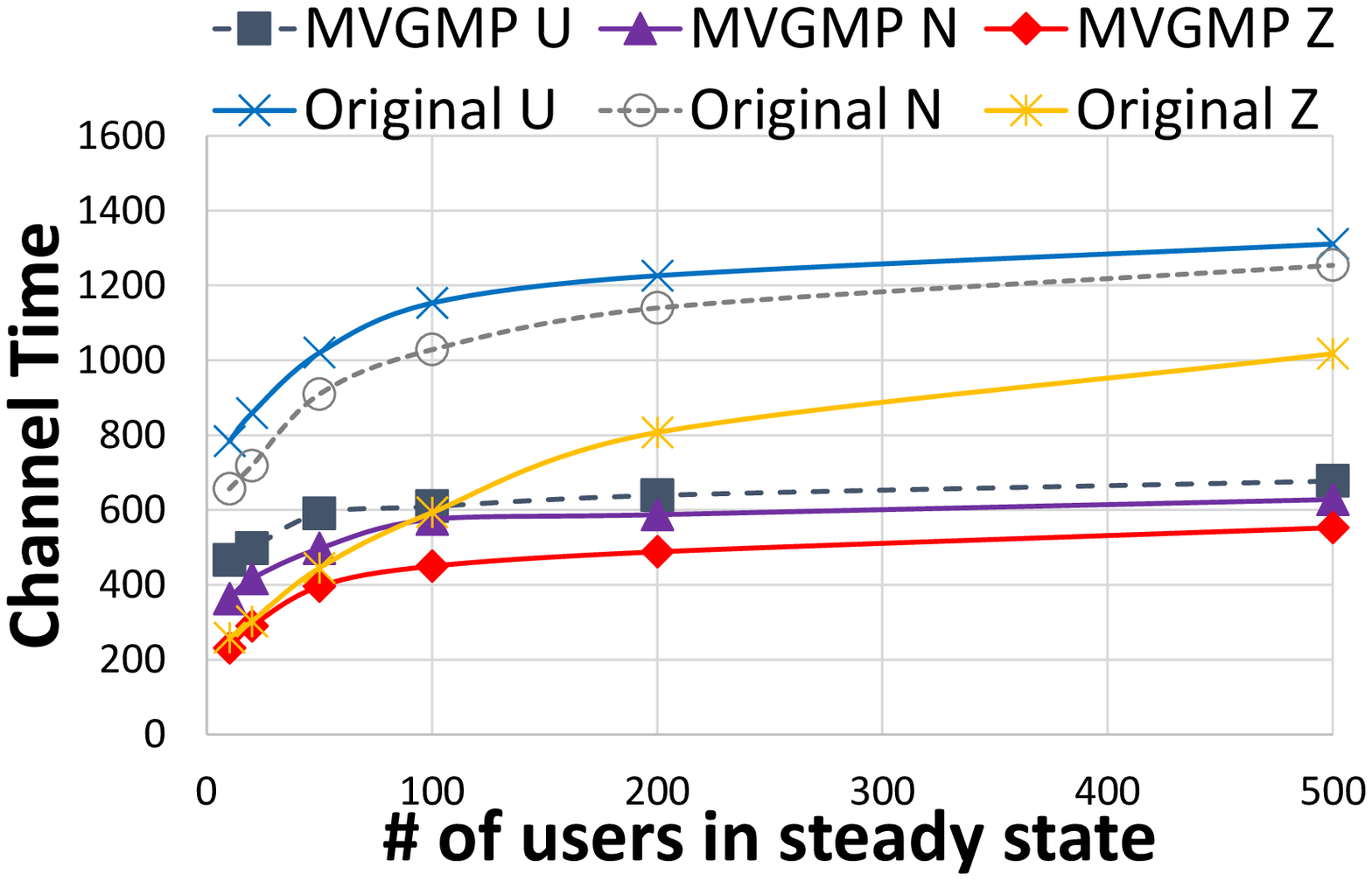}}

\subfloat[Fig. 4: Network Load]{\includegraphics[width=2.2in]{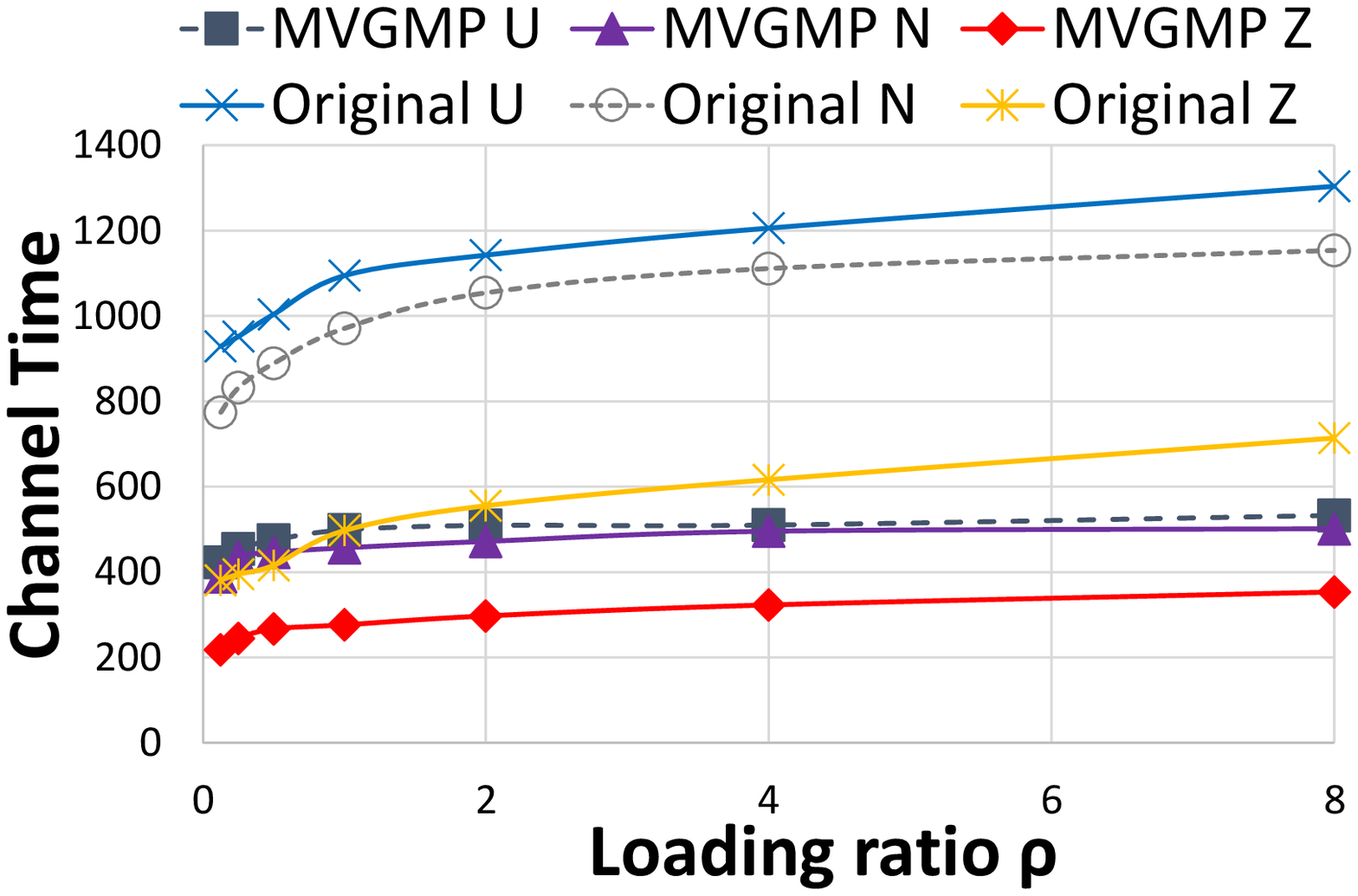}}
\hfil
\subfloat[Fig. 5: View Failure Probability]{\includegraphics[width=2.2in]{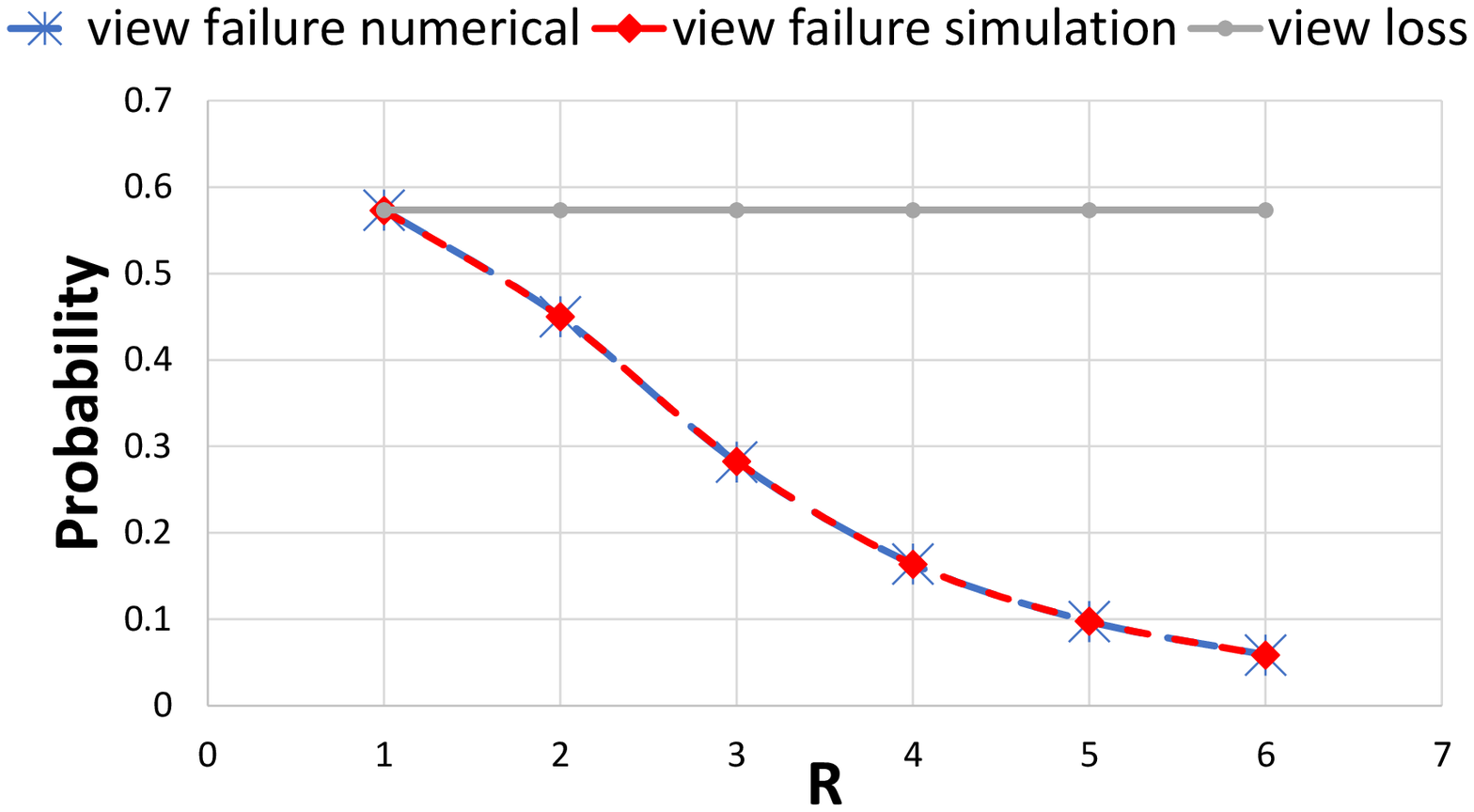}}
\hfil
\subfloat[Fig. 6: Ratio of Successfully Received Views]{\includegraphics[width=2.2in]{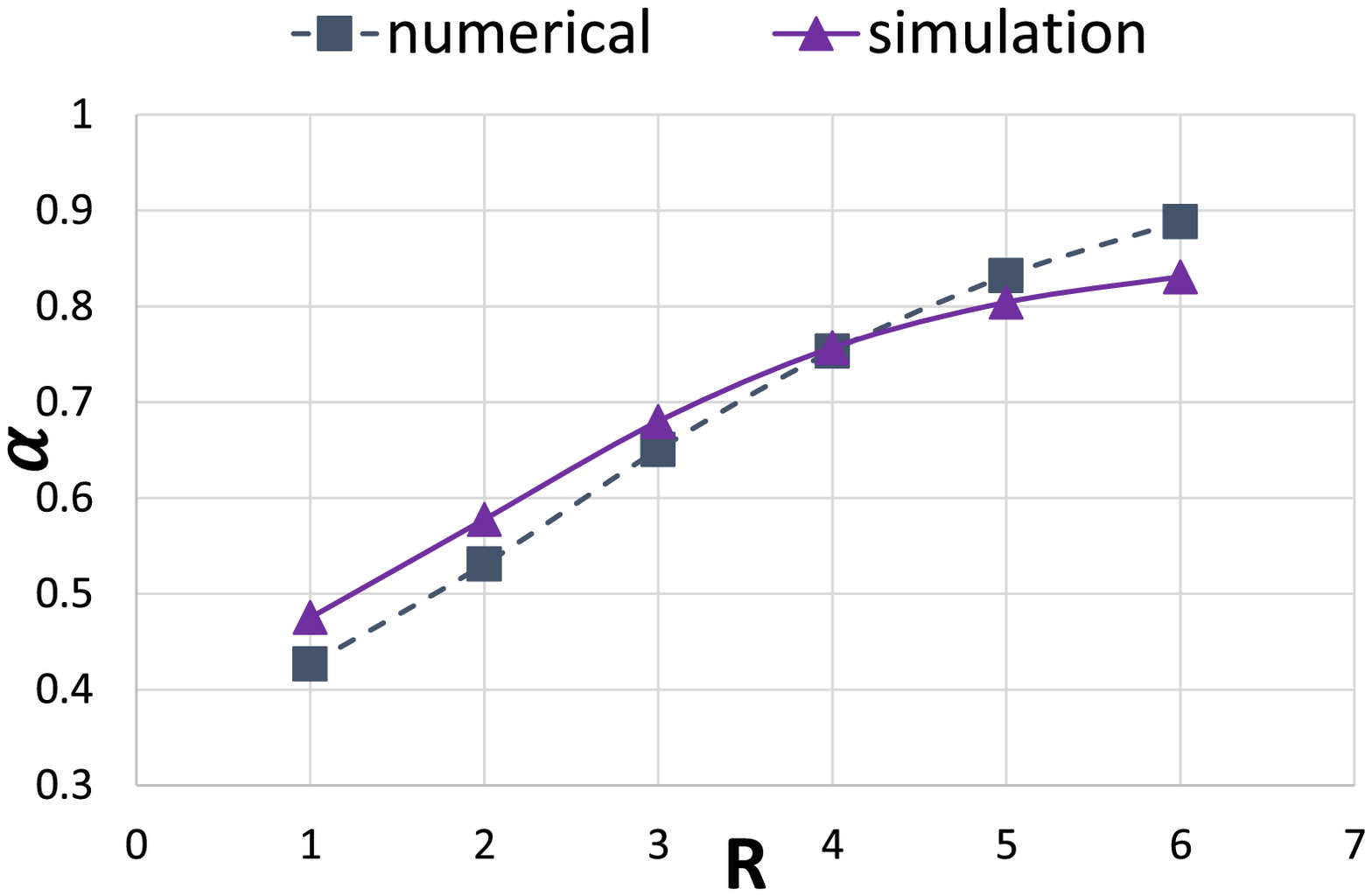}}
\caption*{}
\end{figure*}
\subsection{Impact of User Preferences}
From Fig. 1 to Fig. 4. the results clearly show that Uniform distribution requires 
the most channel time compared with Zipf and Normal distributions. This is because 
in Zipf and Normal distributions, users prefer a sequence of hot views, and those 
views thus have a greater chance to be synthesized by nearby views with DIBR.

\subsection{Analytical Result}
Fig. 5 and Fig. 6 compare the simulation results from NS3 and the analytical results 
of Theorem 1 and Theorem 2 for the Uniform distribution, where each user subscribes to 
each view with a probability of 0.8. The results reveal that the discrepancy among 
the simulation and analysis is very small. Most importantly, $\alpha $ increases for a 
larger $R$ since each user can synthesize and acquire a desired view from more
candidate right and left views when the desired view is lost during the transmissions.

\section{Conclusion}

With the emergence of naked-eye mobile devices, this paper proposes to incorporate DIBR 
for multi-view 3D video multicast in WiFi networks. We first investigated the merits of 
view protection via DIBR and showed that the view failure probability is much smaller than 
the view loss probability, while the multi-view subscription for each client was also 
studied. Thereafter, we proposed the Multi-View Group Management Protocol (MVGMP) to 
handle the dynamic joining and leaving for a 3D video stream and the change of the 
desired view for a client. The simulation results demonstrated that our protocol 
effectively reduces the bandwidth consumption and increases the probability for each client 
to successfully playback the desired view in a multi-view 3D video.

\section{CoRR}
To investigate the case where user subscribes a consecutive sequence of views, we adopt the following setting. User subscribes views according to a Zipf distribution, which means the $k$th view is subscribed with probability $\frac{c}{(k\textrm{ mod }m)^s}$ independently to other views. Figure$7$ depicts this scenario using $m=5$ as an example.

Following theorem serves as a counterpart of theorem $2$ in our main article.
\begin{thm}
In the consecutive view subscription scenario as described above, the ratio $\widetilde{\alpha}$ of expected number of views that can be received or synthesized to the number of total subscribed views tends to
\begin{align}
&p\,\Bigg\{\sum_{j=1}^m\sum_{x=1}^R\Bigg[\Bigg(\sum_{l=1}^{m-j}\frac{c}{(j+l)^s}+\Bigg(\sum_{t=1}^m\frac{c}{t^s}\Bigg)
\frac{x-(m-j)}{m}  \notag \\
&+\sum_{l=1}^{[x-(m-j)] \textrm{mod }m}\frac{c}{l^s}\Bigg)p(1-p)^{x-1}\Bigg]\Bigg\}
\Bigg/{\sum_{l=1}^m\frac{c}{l^s}}  \notag  
\end{align}
as $|\mathcal{K}_i|\rightarrow\infty$, where $p=1-\prod_{c\in C_{i},r\in
D_i}\sum_{n}p_{c,r}^{\text{AP}}(n)p_{i,c,r}^{n}$
\end{thm}

\textbf{Proof:} We follow a similar arguments in our main article, which derives the theorem by reward theory. This time, however, we should use a generalized reward process, the Markov reward process. Let $T_n$ denote the index of the n-th successfully received view, and $G_n$ denote the state of the embedded Markov chain, which represents the "position" of the n-th renewal cycle. An example of this definition is represented in figure. 7, in which the states of the first, second and the third cycles are $1,1,4$ respectively.

The transition probability of $G_n$ is
\begin{numcases}{p_{ij}}
   \frac{p(1-p)^{j-i-1}}{1-(1-p)^m}, & $1\leq i<j\leq m$ \nonumber\\
   \frac{p(1-p)^{m-i+j-1}}{1-(1-p)^m}, &  $1\leq j\leq i \leq m$
\end{numcases}%
since, for example $1\leq i<j\leq m$, the position change from $i$ to $j$ occurs if and only
if there are $j-i$ plus a multiple of $m$ views between the nearest two successfully received views, which means
\begin{small}
\begin{equation}
p_{ij}=p(1-p)^{j-i-1}+p(1-p)^{j-i-1+m}+p(1-p)^{j-i-1+2m}+\cdots\nonumber
\end{equation}
\end{small}

The $\{(G_n,T_n),n=1,2,3,\dots\}$ so defined is then a Markov renewal process.

\begin{figure}[!t]
\captionsetup{labelformat=empty}
\includegraphics[width=3in, angle=270]{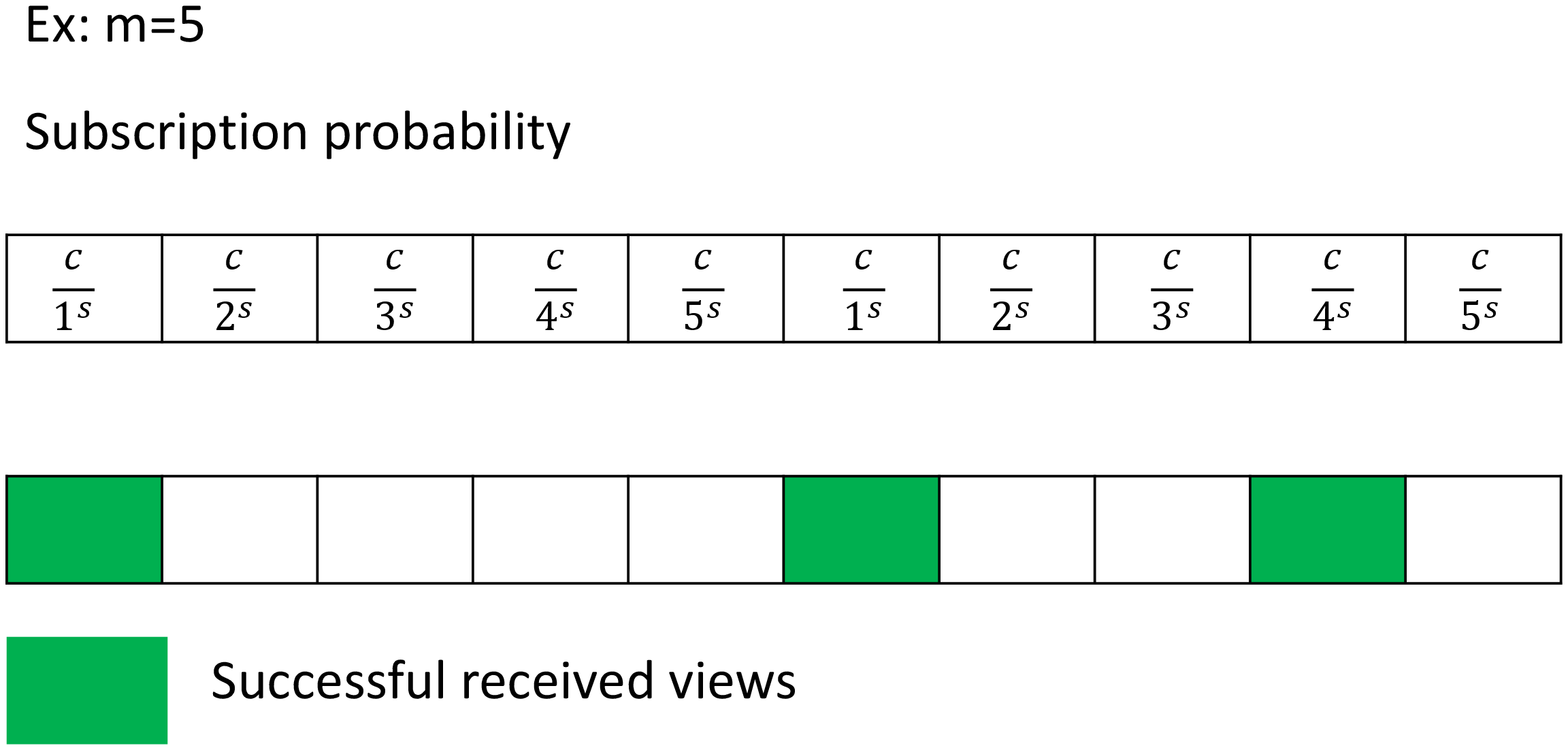}
\vspace{-3cm}
\caption{Fig. 7: Example of consecutive view subscription scenario}
\end{figure}


If we define the reward function of the process as $\rho (j,x)=$
\begin{small}
\begin{numcases}{}
  \sum_{l=1}^x\mathbf{1(\textrm{view in the $l$ position has been subscribed})}, &$x \leq R$ \nonumber\\
  0, &$x> R$ \nonumber
\end{numcases}%
\end{small}

then
\begin{align}
Z_{\rho}=\sum_{n:T_{n+1}<t}\rho(G_n,T_{n+1}-T_n)+\rho(G(t),X(t))
\end{align}
is a Markov reward process, where $X(t)$ is the age process and $G(t)$ be the semi-Markov process associated with our interested Markob renewal process $\{(G_n,T_n),n=1,2,3,\dots\}$.
The process so defined as the following desired property:
The process just defined has a direct relation to our desired quantity $\widetilde{\alpha}$, which is
\begin{align}
\widetilde{\alpha}=\frac{EZ_{\rho}}{S_t}\nonumber
\end{align}
where $S_t$ is the number of views subscribed by the user.
We now intend to apply the theorem 4.1 in \cite{soltani1998} to the right hand side of the above equation. In the following, we will use the same notations as in the article just mentioned.
\begin{align}
h(j)&=\sum_{x=1}^{\infty}\rho(j,x)\sum_{j=1,2,\dots}P(G_{n+1}=j,T_{n+1}-T_n= x|G_n=i)\nonumber\\
&=\sum_{x=1}^{\infty}\rho(j,x)p(1-p)^{x-1}\nonumber\\
&=\sum_{x=1}^R\Bigg[\Bigg(\sum_{l=1}^{m-j}\frac{c}{(j+l)^s}+\Bigg(\sum_{t=1}^m\frac{c}{t^s}\Bigg)\,
\frac{x-(m-j)}{m}\!\nonumber\\
&+\sum_{l=1}^{[x-(m-j)] \textrm{mod }m}\frac{c}{l^s}\Bigg)p(1-p)^{x-1}\Bigg]\nonumber\\
\end{align}
Observe that the steady state of the chain $G_n$ is uniform distribution, which means
\begin{align}
\pi_i=\frac{1}{m}
\end{align}
Now apply theorem 4.1 in \cite{soltani1998}, we have
\begin{align}
\mathbb{E}Z_{\rho}(t)=pt\sum_{j=1,2,\dots}\pi_jh(j)+o(t)\nonumber
\end{align}
Hence,
\begin{align}
&\frac{\mathbb{E}Z_{\rho}(t)}{S_t}\rightarrow mp\frac{\sum_{j=1,2,\dots}\pi_jh(j)}
{\sum_{l=1}^{m}\frac{c}{l^s}}\nonumber\\
&=p\Bigg\{\sum_{j=1}^m\sum_{x=1}^R\Bigg[\Bigg(\sum_{l=1}^{m-j}\frac{c}{(j+l)^s}+\Bigg(\sum_{t=1}^m\frac{c}{t^s}\Bigg)
\frac{x-(m-j)}{m}\nonumber\\
&+\sum_{l=1}^{[x-(m-j)] \textrm{mod }m}\frac{c}{l^s}\Bigg)p(1-p)^{x-1}\Bigg]\Bigg\}
\Bigg/{\sum_{l=1}^{m}\frac{c}{l^s}}\nonumber
\end{align}

\bibliographystyle{IEEEtran}
\bibliography{IEEEabrv,reference}
\end{document}